
\input phyzelv
\input tables
\input defj

\hsize=16cm
\vsize=23.2cm
\hoffset=0pt \voffset=0.5cm
\def\PRD#1 {{\it Phys. Rev.} {\bf D#1}\ }
 
\newdimen\digitwidth \setbox0\hbox{\rm0}
\digitwidth=\wd0 \catcode`?=\active \def?{\kern\digitwidth}

\def\chapter#1{\par \penalty-300 \vskip\chapterskip
   \spacecheck\chapterminspace
   \chapterreset \leftline{\bf\chapterlabel \ #1}}
\def\section#1{\par \ifnum\the\lastpenalty=30000\else
   \penalty-200\vskip\sectionskip \spacecheck\sectionminspace\fi
   \wlog{\string\section\ \chapterlabel \the\sectionnumber}
   \global\advance\sectionnumber by 1  \noindent
   {\bf\chapterlabel \sectionlabel #1}\par  }

\def\brr{\overline    }

\def\Kef{\ifm{ {K_{e4}} }}
\def\tl{\ifm{ {\theta_l} }}
\def\sp{\ifm{ {s_\pi} }}
\def\tp{\ifm{ {\theta_\pi} }}
\def\sL{\ifm{ {s_l} }}

\parindent=1cm
\baselineskip=16pt
\vbox to 5pt{\vfill}\par

\vbox to 1.6 cm{\fourteenpoint
\vfill
{\kern-7mm\leftline{\bf ACCURACIES OF $K_{e4}$ PARAMETERS AT \DAF}}}

\twelvepoint
\vglue 7pt
\leftline{{\bf  Marc Baillargeon}\foot{On leave from
 Laboratoire de Physique Nucl\'eaire, Universit\'e de Montr\'eal,
C.P.  6128, Succ. A, Montr\'eal, Qu\'ebec, H3C 3J7, Canada.}}
\leftline{Laboratoire  de Physique Th{\'e}orique
ENSLAPP,\foot{URA 14-36 du CNRS, associ{\'e}e {\`a}\ l'E.N.S.
de Lyon, et au L.A.P.P. d'Annecy-le-Vieux}}
\leftline{B.P. 110, F-74941  Annecy-Le-Vieux, Cedex, France}
\leftline{\bf Paula J. Franzini}
\leftline{Paul Scherrer Institute, CH-5232 Villigen PSI, Switzerland}
\vglue 7pt
\vglue 7pt
\leftline{\hskip2cm\bf Abstract}
\vglue 7pt
\parshape=1 2cm 12cm
\noindent
We estimate the experimental accuracies to which the \Kef\ parameters
will be measured at \DAF.  We consider the accuracies obtainable using
 asymmetries, and the maximum
likelihood method.  We find the current determinations of the relevant
parameters will be improved by a factor of five to ten after a year
of running at the anticipated luminosity of \DAF.
\vglue 7pt
\parshape=0

\chapter{Estimating experimental uncertainties: Generalities}
 In this contribution to the \DAF\ physics handbook, we estimate the
accuracies achievable at \DAF\ in measuring the  theoretical
parameters describing the decay $K^+ \rightarrow \pi^+ \pi^- e^+ \nu_e$.
We refer the reader to the contribution of Bijnens, Ecker, and
Gasser\Ref\BEG{J. Bijnens, G. Ecker and J. Gasser,
{\it \DAF\ Physics Handbook}, vol. 1, p. 115, 1992.}
for a detailed discussion of the theory of this decay.
We should note that we refer to this decay as \Kef; we
consider only this particular decay and not the channels with
neutral pions or kaons.

We will consider two of the possible ways
of extracting the parameters of a theory from a set of data.
The first one is the classic technique of asymmetries.
\Kef\ decays have a partial
decay rate ${\rm d}^5\Gamma$ of the form
$${\rm d}^5\Gamma = G^2_F |V_{us}|^2 N(\sp,\sL)J_5(\sp,\sL,\tp,\tl,\phi)
{\rm d}\sp {\rm d}\sL {\rm d}(\cos\tp) {\rm d}(\cos\tl){\rm d}\phi.
\eqn\eqviib$$
Here \sp\ and \sL\ are the effective mass squareds of the dipion and
dilepton systems; \tp\ is the angle of the $\pi^+$ in the dipion
c.m. frame, with respect to the dipion direction in the $K$ c.m. frame,
\tl\ is the analogous angle for the $e^+$, and $\phi$ is the
angle between the pion plane and the lepton plane in the $K$ c.m. frame.
$N$ is a kinematical factor depending only on \sp\ and \sL\
and none of the theoretical parameters.
The quantity $J_5$ can be written as an expansion in sines and cosines
of $\phi$ and \tl\ multiplying nine
intensities  $I_i$, which can in turn be written in
terms of the three remaining kinematical variables, and three
form factors, $F$, $G$, and $H$.  The explicit dependences may be
found in Ref. \BEG.
As can be seen in the contribution of Colangelo, Knecht and
Stern,\Ref\CKS{G. Colangelo, M. Knecht and J. Stern, in this
Handbook} the
tangent of the phase shift, $\tan(\delta_0-\delta_1)$,
can be neatly extracted from the ratio of the intensity
functions, $\brr{I_7}/2\brr{I_4}$, or, equivalently,
$\tilde{I_7}/2\tilde{I_4}$, where by the tilde we denote
intensities integrated over $s_\pi$ and $s_l$ as well
as over \tp.
   All of the $\tilde{I_i}$, in
turn, can be written as asymmetries; in particular we have
$${8\over 3\pi} \tilde{I_4} = \left(\int_0^{\pi/2} \int_{3\pi/2}^{5\pi/2}
-\int_0^{\pi/2} \int_{\pi/2}^{3\pi/2} -
 \int_{\pi/2}^{\pi} \int_{3\pi/2}^{5\pi/2} +
 \int_{\pi/2}^{\pi} \int_{\pi/2}^{3\pi/2}\, \right)
\sin\tl{\rm d}{\tl}{\rm d}{\phi} \int J_5 {\rm d} \sp
 {\rm d} \sL  {\rm d} \tp \eqn\eqi$$
and
$$ \tilde{I_7} = \left(\int_0^{\pi} - \int_{\pi}^{2\pi}\,  \right)
{\rm d}{\phi} \int J_5 {\rm d} \sp
 {\rm d} \sL  {\rm d} \cos\tp{\rm d} \cos\tp.
\eqn\eqii$$
The asymmetry presents a transparent, elegant and quick
(computationally) way to determine a parameter.

The second, and main, method we consider is that of the maximum
likelihood, which we shall refer to as the MLM.\Ref\mlmref{For more
details see for example P. Franzini
in the {\it \DAF\ Physics Handbook}, 1st ed.,
vol. 1, p. 15, 1992.}
Let $\vec{x}$ be the vector of phase space variables specifying
an event in an experiment, and $f(\vec{x};\vec{p})$ be the
probability distribution function predicted by a theory.
 $\vec{p}$ is the set of parameters in the theory, to
be determined experimentally.  The probability of observing
an event at $\vec{x}$ in the interval ${\rm d}^n x$ is
$f(\vec{x};\vec{p}){\rm d}^n x$.  The function $f$ is normalized
to 1 over the whole $\vec{x}$ interval in which $\vec{x}$
is physical.  In particular, if cuts are imposed on the phase space,
$f$ must be normalized over the reduced phase space; we must
also be certain that the normalization is maintained even when
the parameters are varied from their central value.

The likelihood,
or joint probability distribution, of an experiment yielding
$N$ events, each
 specified by a set of phase space variables $\vec{x_i}$, is
then defined as
$$ {\cal L} = \prod_{i=1}^{N} f(\vec{x_i};\vec{p}).
\eqn\eqiii$$
The best estimate for $\vec{p}$ is then simply the value
$\brr{p}$ which maximizes ${\cal L}$ (or equivalently, of
$W=\log {\cal L}$, which is easier to compute).
It can be shown that the error matrix, in general non-diagonal
for correlated parameters, is
$$ \brr{ (p_i - \brr{p}_i) (p_j - \brr{p}_j) } = {1\over N} \int
\left[ {1\over f} \left( {\partial f(\vec{x};\vec{p})\over\partial p_i}
 {\partial f(\vec{x};\vec{p})\over\partial p_j} \right) {\rm d}^n x
\right] ^{-1} .
\eqn\eqiv$$
Here the $[\ ]^{-1}$ denote matrix inversion.  This integral
is, in general, not possible to compute analytically, but is
easily evaluated numerically.
If we start with a non-normalized
probability function ${\cal P}$,
then the following equation is useful:
$$ \int {1\over f}  {\partial f\over\partial p_i}
  {\partial f\over\partial p_j} {\rm d}^n x =
-{1\over a^2}  {\partial a\over\partial p_i}
  {\partial a\over\partial p_j}
+ \int {1\over a{\cal P}}  {\partial {\cal P}\over\partial p_i}
  {\partial {\cal P}\over\partial p_j} {\rm d}^n x,
\eqn\eqv$$
where $a=\int {\cal P} {\rm d}^n x$.

While this method is more complicated and time-consuming
computationally, it has the decided advantage that it
yields the absolute best possible determination of any
given set of parameters.  In today's age of fast computers,
it is thus the method of choice.  It also allows one to
determine {\it any} parameters chosen, while the asymmetry
is only good for certain parameters.  Finally, the
experimental {\it uncertainties} in the parameters
can be determined reasonably easily, without need for an
actual simulation of the determination of the parameters
themselves using this method.

It is simple to see that the MLM should give a better
determination of the parameters than an asymmetry:
while the asymmetry only uses the information of
whether an event is in one or another half of phase
space, the MLM benefits from the information of the
precise position of the event.  For example, consider
a process specified by the probability distribution
$$f(\theta;a) = (1+a\cos\theta)/2.  \eqn\eqvi$$
The parameter $a$ can be determined by the ratio
$(N_1-N_2)/(N_1+N_2)= a/2$ where $N_1$ is the number of events
with $\theta$ between 0 and $\pi/2$, and
$N_2$ is the number of events
with $\theta$ between $\pi/2$ and $\pi$.
For $a$ small compared to 1, the error on $a$ in this
determination is thus $2/\sqrt{N}$ where $N$ is the total
number of events.
For the MLM, the integral in eq.~\eqiv\ is
$$\lim_{a\To 0} \int_{-1}^{1} {x^2\over 2(1+ax)} {\rm d} x
={1\over 3} \eqn\eqvii$$
and the error on $a$ in this determination is thus
$\sqrt{3/N}$.  While this improvement may not seem
very impressive, larger improvements may be expected
as the parametrization becomes more complicated.
Using the MLM directly on the parameter that interests us,
 we are sure to use all possible information,
and to take into account the effect of all possible
cancelling uncertainties.
We observe an improvement as above, of about fifteen
percent, in the relative error $\Delta \tilde I_7/\tilde I_7$
when we go from the asymmetry method to the MLM.
However, the relative error
on $\delta = \delta_0-\delta_1$ when $\delta$ is determined
{\it directly} via the MLM is two-thirds of its error when
determined {\it indirectly} by the ratio
$\tilde{I_7}/2\tilde{I_4}$, if these $\tilde I_i$
are determined using the MLM, and their errors are then combined
as uncorrelated.  This appears to be due to the information
lost in integrating the $\tilde{I_i}$ over \sp, \sL, and \tl\
before using the MLM.

\chapter{Estimating experimental uncertainties in \Kef}
  For our first estimates, we consider only the
first order terms in a partial wave expansion of the form
factors $F$, $G$ and $H$, i.e., we take
$$F = f_s e^{i\delta_0} \ \ \ \ \ G= g e^{i\delta_1}
\ \ \ \ \ H=h e^{i\delta_1}. \eqn\eqviii$$
This is consistent with the parametrization used by
Pais and Treiman,\Ref\PT{A. Pais and S. B. Treiman,
Phys. Rev. {\bf 168} (1968) 1858} and Rosselet {\it et
al.}\Ref\Ross{L. Rosselet {\it et al.}, Phys. Rev.
{\bf D15} (1976) 574.}  They consider higher order
terms, but the coefficients of these terms are found to
be consistent with zero by the experiment of Rosselet
{\it et al.},\refmark\Ross  so we do not consider
further terms in our initial estimates.

All the $\phi$ and \tl\ dependence in the problem is
contained in the expression of $J_5$ as a function of
these variables and the $I_i$; for this reason these
two variables are referred to as ``trivial.''  \tp\
appears only in the equations
for the $I_i$ in terms of $F$, $G$, and $H$ (and the phase space
expansion, should we consider higher order terms).
This leaves then $f_s$, $g$ and $h$ with a possible
dependence on \sp\ and \sL.  The $\pi\pi$ phase
shifts $\delta_0$ and $\delta_1$ (which at this order
appear only in the combination $\delta=\delta_0-\delta_1$)
depend only on \sp.  For the moment we parametrize
$f_s$, $g$ and $h$ by an expression of the form
$$ y(\sp,\sL) = y_0 (1 + \lambda q^2) \eqn\eqix$$
where $q^2 = (\sp - 4 m_\pi^2)/4m^2_\pi$, and $y$ stands
for $f$, $g$, or $h$.   We take
the slope $\lambda$ to be the same for $f_s$, $g$
and $h$, and no slope in \sL\ at this stage, again
consistent with Ref. \Ross.  For the dependence of $\delta$
on \sp\ we will consider average values in a set of
5 bins in \sp, and consider parametrizations of $\delta$
in a later section.

One last important detail remains to be mentioned.  The
MLM (or asymmetries, for that matter) says nothing about
an {\it overall factor} in the intensity, as we require the
probability density to be normalized to one.  Thus we
divide out $f_{s0}$ ($f_0$ for short) from the amplitude, as it
is the parameter with the most effect on the integrated intensity.
Wherever $g_0$ and $h_0$ appear, they are divided by $f_0$,
so we replace them by new parameters
$g'_0=g_0/f_0$ and $h'_0=h_0/f_0$ ($\lambda$
and $\delta$ are unaffected).  We then apply the MLM
to the set of parameters $\delta$, $g'_0$, $h'_0$ and $\lambda$,
and obtain the correlation matrix
$$({\rm d} p {\rm d}p)_{ij} = {1\over N}
\left(\matrix{ 6.5 & 0.8 & -0.7 & -0.1 \cr
0.8 & 2.8 & -0.6 & -0.8 \cr -0.7 & -0.6 & 187 & 0.4 \cr
-0.1 & -0.8 & 0.4 & 3.8 \cr} \right) . \eqn\eqx$$
The diagonal entries of this matrix are  variances of the
four parameters, where $N$ is the number of
events.  The off-diagonal elements represent correlations between
the parameters; in this case they are small, but they can be significant,
depending on the parametrization used.  We
 do not report the full correlation matrix for each parametrization
 in this paper, but they are available from our
programs if needed for further calculations.
They can not be  neglected in general if one wants to calculate
functions of the parameters we use, and propagate the errors
correctly.  In the end of this paper, we consider the determination
of some highly correlated parameters.

We then extract the error on $f_0$ from the equation
$$ \Gamma_\Kef = C f^2_0 \int {\cal P} {\rm d}^n x, \eqn\eqxi$$
where ${\cal P}$ is the unnormalized probability function
inputted to the MLM calculation, and $C$ represents all the
constant factors (masses, two's, $\pi$'s) needed to complete
the equation.  The relative error $\Delta\Gamma/\Gamma$ is
given by $\approx 1/\sqrt{N}$ in an experiment like
KLOE,\Ref\KLOE{See for example {\it The KLOE detector
technical proposal}, by the KLOE collaboration, A. Aloisio
et al., LNF-93/002.}
where the statistical error will be dominant.
The relative error on the integral ($a$) is given by the matrix
product
$$ {\Delta a \over a} = \sqrt{{1\over a^2}
{\partial a \over \partial p_i} ({\rm d} p {\rm d} p)_{ij}
{\partial a \over \partial p_j}}. \eqn\eqxii$$
Combining this error in quadrature with the statistical error on
$\Gamma$, we obtain the error we quote for $f_0$; combining
the error on $f_0$
in quadrature with the errors on $g'_0$ and $h'_0$,
we obtain the errors we quote for $g_0$ and $h_0$.

In Table 1 we display the results of this calculation.
The central values (our input) are those found by the previous
experiment.\refmark\Ross
We have used the program VEGAS\Ref\VEG{G. P. Lepage,
CLNS-80/447.}
to do the necessary integrals in five-dimensional phase space.
The normalization of the probability distribution is ensured
automatically by the program, and the necessary derivatives
also computed numerically.
 Estimated errors are shown for
$N=30000$ events, the statistics of the previous experiment)
and $N=300000$ events, the anticipated statistics\refmark\BEG
in one ``year''$\equiv 10^7$ seconds
of running with ${\cal L}= 5 \times 10^{32}
{\rm cm}^2 {\rm s}^{-1}$.
All errors in this paper, unless otherwise noted, are statistical
errors and can be simply scaled by $1/\sqrt{N}$ for different
numbers of events.  As a general rule, also, the {\it fractional}
error, on parameters such as $f_0$, $g_0$ and $h_0$, is roughly
independent of the central value taken for them, while the
{\it absolute} errors on parameters such as $\delta$ and $\lambda$
remain constant.  The last line of Table 1 shows the errors on
these parameters found by Rosselet {\it et al.}.
We do not quote the error on $\delta$ because an error on
$\delta$ averaged over the whole of phase space is not very
meaningful.  The errors shown are independent of the central value
of $\delta$ used.

\midinsert
{\bf Table 1.}  Central values  and estimated errors
for $f_0$, $g_0$, $h_0$ and $\lambda$
\vskip 9pt
\begintable
      \ \             | $f_0$  | $g_0$ | $h_0$ | $\lambda$  \cr
   Central values  |  5.59  | 4.77 | $-$2.68 | 0.08       \cr
 Errors ($N=30000$) | 0.029 | 0.059 | 0.44 | 0.011   \cr
 Errors ($N=300000$) | 0.009 | 0.019 | 0.14 | 0.004   \cr
Errors (Rosselet) | 0.14 | 0.27 | 0.68 | 0.02
     \endtable
\endinsert

At this point, before going on to further discuss errors in KLOE, it is
necessary to say some words about  why our estimated errors at
``Rosselet statistics'' are so different from those that Rosselet quotes.
The errors given above are purely statistical, but apply to a
``perfect'' detector, i.e., one which covers the whole of phase space with
unity efficiency everywhere.
  This is close
to true for KLOE; we will attempt to illustrate this later in this
paper, and will describe a more rigorous demonstration in
a future paper.  However, Rosselet's detector was far from ``perfect.''
In the error we have quoted for $f_0$, the errors from $\Gamma_\Kef$
and the normalization $a$ contribute about equally; the first is
about $0.6\%$ and the second about $0.9\%$.  Rosselet, however,
quotes a relative error $\Delta \Gamma_\Kef / \Gamma_\Kef$ of
$4.5\%$, which completely accounts for their large error on $f_0$.
Their fixed target experiment had a $10\%$ overall efficiency for \Kef,
and a highly variable efficiency as well, varying, for example,
smoothly from $>95\%$ in a very small portion of phase space with
large \sL\ and small \sp, to near zero at large \sp\ and small \sL.
KLOE is in contrast a hermetic detector, operating at a $e^+ e^-$
collider running at the $\phi$ resonance, producing self-tagging
low momentum $K^\pm$ pairs.  It will have a uniform near-$100\%$
efficiency over all of phase space, minus a few percent of phase
space that will be cleanly cut and discarded.\Ref\PFp{P. Franzini,
private communication.}

The next step in our analysis was to drop the slope parameter
$\lambda$ and determine the errors on the parameters in five
bins in \sp, chosen so as to have equal numbers of events.  Such
an analysis with real data would have the advantages of studying
the \sp\ dependence in a more parametrization
independent way.  If, however,
the \sp\ dependence is correctly given by eq. \eqix, this method
will not determine $\lambda$ as accurately, so in general both
types of approach are necessary.  For our purposes, displaying the
error in bins is also important to illustrate the possible accuracies
with which $\delta(s_\pi)$ may be measured, before we implement a possible
parametrization of $\delta$.  In Table 2, we give the estimated
errors, taking an average of $y_0 (1 + \lambda q^2)$ in each bin
as our inputs for $y=f,$ $g,$ and $h$, with Rosselet values for
the $y_0$ and $\lambda$.  The errors on $\delta$
are essentially independent of the inputs of its central value.
In the last line, for comparison, we display the errors
on $\delta$ as measured by Rosselet {\it et al.}  The improvement
is not as drastic as that of $f_0$ was, but is nonetheless a factor of
1.5 to 2.  This should be further multiplied by a factor of
$\sqrt{10}$ to $\sqrt{20}$ per \DAF\ running year.
The accuracy on $f$ in bins is even better than we might have
expected from the error on $f_0$ multiplied by $\sqrt{5}$.  This
is because the error on $\lambda$ gives most of the contribution
to the error on the normalization $a$, and thus a significant
contribution to the error on $f_0$.

\midinsert
{\bf Table 2.}  Estimated errors in five bins of 6000 events each.
\vskip 9pt
\begintable
$\sqrt{s_\pi}$ (GeV) | $0.279-0.3$ | $0.3-0.316$ | $0.316-0.334$ |
$0.334 - 0.357$ | $0.357 - 0.494$ \cr
      $f$      | 0.037  | 0.039 | 0.041 | 0.043 | 0.047 \cr
   $g$  |  0.195 | 0.142 | 0.124 | 0.116 | 0.112 \cr
  $h$   |  1.46 | 1.04 | 0.93 | 0.90 | 1.00 \cr
$\delta$ | 0.062 | 0.041 | 0.034 | 0.029 | 0.025 \cr
$\delta$ (Rosselet) | 0.13 | 0.07 | 0.05 | 0.04 | 0.04
     \endtable
\endinsert

We have examined in some detail the question of what accuracy
$\delta$ can be measured to.  We have first of all determined
that while $\delta$ appears in $I_1$, $I_2$, $I_4$, $I_5$,
$I_7$, and $I_8$, it is only the dependence of $I_7$ that
gives us the above accuracy on $\delta$.  This can be seen
by replacing $\delta$ in all the $I_i$, except $I_7$, by
a dummy variable $\delta_0$, set equal to the central value
of $\delta$.  When we proceed to apply the MLM to the new
probability function, we find the same error on $\delta$
as before, within a few percent.  If, however, we apply
the MLM to the $\tilde I_i$ as parameters in their own right
(we cannot use the $\brr I_i$ as parameters, because they are
functions of the phase space variables)  and then take the
ratio to determine $\delta$, we find that the error on $\delta$
increases by $50\%$. (If we use the asymmetry method to
determine the $I_i$, the error increases another $15\%$.)
 We have not taken care to cancel
correlated errors in $\tilde I_4$ and $\tilde I_7$,
but we have checked that
the correlated parts of the errors are small relative to the
uncorrelated parts.
So, this $50\%$ increase appears to be mainly due to the information
lost in integrating the $I_i$ over three out of five of the
phase space variables before applying the MLM.  Equivalently,
the better error on $\delta$ can be attributed to applying
a more detailed parametrization (therefore more information)
from the beginning of the calculation.

Nonetheless, it may be interesting to determine the $\tilde I_i$ and
their errors
 as a parameterization independent way to present
the data.  We have estimated that the combination $I_1 -I_2/3$
can be determined to $0.4\%$ in five bins of 60,000 events each.
The other $I_i$ can be determined with absolute errors of
one to two times this error.

\chapter{Other experimental uncertainties}
So far we have estimated the statistical experimental uncertainties
in a perfect detector.  In this section we would like to say a few
words about other experimental uncertainties.
Errors (other than those already estimated) in a detector like KLOE,
where the efficiency is essentially $100\%$ outside of a small
region of phase space to be removed by cuts, fall into two categories:
those associated with cuts, and those associated with resolution.
We intend to write an event generator, that would allow us to
impose a proper cut
on  true kinematical variables such as the angles of the particles
themselves, or their momenta,
 but for the moment are only
calculating the MLM error estimating integral
over \sL, \sp, \tl, \tp\ and $\phi$.
We expect that\refmark\PFp
appropriate cuts will be applied
to reject charged particles that have momenta
less than 20 MeV, and those that are within $9\deg$ of being
parallel to the beam pipe (because they will not cross enough
wires).  We expect that both of these cuts will (from the point of
view of our problem) reject events fairly randomly, and thus only
have the effect of reducing the number of events by a couple of
percent, having a negligible effect on our error estimates.
Just to check, we have considered cutting $s_\pi$ in the range
$4m_\pi^2$ to $4m_\pi^2+4 * (20 {\rm MeV})^2$, and $s_l$
in the range $0$ to $(20\ {\rm MeV})^2$.  The \sp\ cut has negligible
effect and the \sL\ cut increases our errors by about 10 percent
or less.  Eventually we would also like to investigate the
effect of inaccurate cuts (for example a cut that is believed to
be at 20 MeV and is really at 15 MeV), which can be an important
source of systematic errors.

Smearing, also known as convolution with a resolution function,
can be imposed on our variables with more confidence, being a random
effect.  We assume a gaussian resolution function, i.e., we
replace the probability function $f(x;p)$ by a
new function $g(x;p)$
$$\eqalign{g(x;p)=&
\int_{x_0}^{x_1}
 f(x'';p) {e^{-(x''-x)^2/2 \sigma^2} \over \sigma
\sqrt{2\pi} } {\rm d} x'' \approx\int_{-\infty}^\infty
 f(x-x';p) {e^{-x'^2/2 \sigma^2} \over \sigma
\sqrt{2\pi} } {\rm d} x' \cr= &
\int_{-\infty}^\infty\!\!\int_{-\infty}^\infty
f(x-x';p) e^{-x'^2/2 \sigma^2} e^{-y^2/2 \sigma^2}
{1\over2\pi \sigma^2} {\rm d} x' {\rm d} y=
\int_0^1\!\!\int_0^1 f(x-x';p)  {\rm d} a {\rm d} b \cr}
\eqn\eqxiii$$
where $y$ is a dummy variable introduced to make the gaussian
integrable, $a = e^{-({x'}^2+y^2)/2\sigma^2}$ and $\tan(2\pi b) = x'/y$.
$x_0$ and $x_1$ are the lower and upper bounds on $x$;
$x_1-x_0$ is assumed to be large compared to $\sigma$.
Thus we have
$$g(x;p) = \int_0^1\!\!\int_0^1 f
\left(x-\sigma\sqrt{-2\log a}\, \sin(2\pi b);p \right)
{\rm d} a {\rm d} b, \eqn\eqxiv$$
showing the equivalence of convolution with a
gaussian and integrating after a gaussian
{\it smearing} of
the independent variable. Note that the quantity
$\sqrt{-2\log a}\, \sin(2\pi b)$, for $a$ and $b$ uniformly distributed
in the interval $0$ to $1$, is gaussian distributed with variance 1.

KLOE expects a resolution of about half a
percent to one percent in angles and momenta.  We make the assumption
that smearing by one percent (of $2\pi$ or $\pi$ in the case of
our angles) in our five variables is a reasonable and generous
approximation
to smearing in the actual kinematic variables.  We observe negligible
effect on our errors; in fact smearing of as much as
$5\%$ of the maximum
of our variables has no effect except in $s_\pi$, where it yields
a ten percent (fractional!)
increase in our errors.  This is not surprising.
If one {\it knows} one's resolution function,  one can compensate
for it, and the accuracy is unaffected, at least if the scale of the
smearing is small compared to the scale of the effect (this scale
 in our case is of the order of
the whole of phase space).  If however
there are {\it unknown} parts to the resolution function, i.e.,
{\it systematic} errors, these can result in systematic errors
in the result, that the MLM error estimating technique above will
never find, simply because we have no way to input an {\it unknown}
error.   These errors could in principle be estimated by applying
the MLM itself to simulated data, but this is beyond the scope of
our investigation.  The accurate estimation of systematic errors
is at any rate something that will have to be done by KLOE.
Meanwhile, however, we have made one attempt
to estimate the effect of such systematic
errors by returning to the determination of $\delta$ from the
asymmetry method.  Here we can easily calculate what
$\delta$ we would ``measure'' from an imaginary set of data
described by a set of input parameters;  normally of course, we
get back the value of $\delta$ we input.   If, however, we replace
the probability distribution $f(x)$ by a convoluted distribution
$g(x)$, then,
by the shift in the recovered $\delta$, we can see what would
be the effect of a resolution function that we did not know about,
and therefore did not compensate for.

We have first verified that gaussian smearing still has no
significant effect, which means that it is not
neccessary to know the exact form of the resolution
function, as long as it is symmetric and not too broad
in the appropriate scale.
We have next examined the effect of an actual shift in each
of our phase space variables.  The effect is tiny compared
to the statistical errors for shifts in \tl, \tp, \sL\
and \sp.  For a $1\%$ shift in $\phi$, $\delta$ goes down
by 0.01, or about half of the statistical error with
300,000 \Kef\ events.  We remark that a $1\%$ systematic
shift in $\phi$ is already so large as to be inconcievable,
as $\phi$ is a difference of observable angles.

\chapter{Other parametrizations}
In this section we would like to present results
for some extensions of the minimal parametrization we have used
so far:  (1) we include a $p$-wave term in $F$, and simultaneously
adopt a notation consistent with the partial wave expansion of
Ref. \BEG; (2) we consider an $s_l$ dependence in $f$, $g$, etc.;
(3) we consider two parametrizations of $\delta$ in terms
of $s$-wave scattering length $a_0^0$ to
give a first indication on what errors can be expected on this
parameter.

 $F$ thus becomes
$$F = f_s e^{i\delta_0} + f_p e^{i\delta_1} \cos\tp
- {\sigma_\pi \, P\cdot L \over X}\cos\tp g e^{i\delta_1},
 \eqn\eqxv$$
where $\sigma_\pi$, $P\cdot L$ and $X$ are defined in Ref. \BEG.
Our previous parametrization is thus equivalent to this one
for $f_p = \sigma_\pi\, P\cdot L \, g/X$.
For the $s_l$ dependence, chiral perturbation theory to one loop
predicts\Ref\GCp{G. Colangelo, private communication.}
the same slope for \sp\ and \sL, or, more precisely,
$$ y(\sp,\sL) = y_0 \left(1 + \lambda
\left({s_\pi -4m_\pi^2-\sL\over 4m_\pi^2}\right)\right).
\eqn\eqxvi$$
To be more model-independent, and to have a separate  determination
of the \sp\ and \sL\ slopes, we choose the parametrization
$$ y(\sp,\sL) = y_0 \left(1 + \lambda q^2\right)
\left(1- \lambda_2 \sL/4m_\pi^2 \right).
 \eqn\eqxvii$$
We then find, for 300000 events,
$$
\Delta f_{s0}= 0.014 \ \ \ \ \Delta g_0 = 0.038
\ \ \ \ \Delta h_0 = 0.14
\ \ \ \ \Delta f_{p0} = 0.014 \eqn\eqxviiia$$
$$\Delta \lambda = 0.004\ \ \ \ \Delta
\lambda_2 = 0.011  \eqn\eqxviiib$$
to be compared with table 1.  We have displayed the errors for
the old parameters as well as the added ones, as some of them
have increased slightly.  We have taken central values
$\lambda_2=0.08$ and $f_{p0}=3.3$ (which maintains the normalization
unchanged); if other parameters are preferred it should be remembered
that the fractional error of $f_{p0}$, and the absolute error of the
slope,  remain essentially constant.

For $\delta$ we have first considered the parametrization
used\Ref\BFP{J. L. Basdevant, C. D. Frogatt and J. L. Petersen,
Nucl. Phys. {\bf B72} (1974) 413.}
by the previous experiment, to compare  our estimated errors
with the ones they determined.  They use
$$\sin 2\delta = 2\sqrt {\sp - 4m_\pi^2 \over s_\pi }\left(
a^0_0 + {b q^2 }\right), \eqn\eqxix$$
where $b=b^0_0-a^1_1$, the difference between the $s$-wave slope
and the $p$-wave scattering length.  There is moreover
a possible relation between $b$ and $a_0^0$
$$b = 0.19 - (a^0_0 - 0.15)^2.  \eqn\eqxx$$
For 30000 events (to compare with Ref. \Ross) we find
$\Delta a^0_0 = 0.029$, to be compared with 0.05 in Ref. \Ross,
if we use both eqs. \eqxix\ and \eqxx.
If we use only eq. \eqxix\ we find $\Delta a^0_0=0.06$ and
$\Delta b = 0.07$, to be compared with 0.11 and 0.16 in
Ref. \Ross.  We have used the central values $a^0_0=0.28$ in
the first case, and $a^0_0=0.31$, $b=0.11$ in the second case,
as found in Ref. \Ross.  So, in one year running at \DAF\ at
a luminosity of $5\times 10^{32}{\rm cm}^2 {\rm s}^{-1}$, we
expect a factor of five improvement in the error on the
$\pi\pi$ scattering length, meaning that the existing discrepancy
between measurements and predictions for $a_0^0$ should be
resolved.

We have also considered  a more recent parametrization, due to
Schenk.\Ref\Schenk{A. Schenk, Nucl. Phys. {\bf B363} (1991) 97.}
He gives
$$\tan \delta^I_l (\sp) = \sqrt{ {\sp-4m^2_\pi \over \sp} }
q^{2l} \left( a^I_l + \tilde{b}^I_l q^2 + c^I_l q^4 \right)
\left( {4m^2_\pi - s^I_l \over \sp - s^I_l} \right)
\eqn\eqxxi$$
where $q^2 = (\sp - 4 m^2_\pi)/4m^2_\pi$ and
$$\tilde{b}^I_l = b^I_l - a^I_l {4m^2_\pi \over s^I_l -4m^2_\pi}
+ \left(a^I_l\right)^3 \delta_{l0}.
\eqn\eqxxii$$
Here $I$ denotes isospin, $l$ angular momentum, and $\delta_{ij}$ is the
Kronecker delta.  We need to calculate $\delta_0\equiv\delta^0_0$
and $\delta_1\equiv\delta^1_1$;  $s_0^0$ and $s_1^1$
(the values at which the phase shifts should pass through $90\deg$)
are the squares of the $\sigma$ and $\rho$ meson masses.
There are far too many parameters here to  be able to
determine them
by \Kef\ measurements alone; to begin with, we fix the higher order
coefficients $c_0^0$ and $c^1_1$ to zero.
The remaining set $a_0^0$, $a_1^1$, $b_0^0$ and $b_1^1$
is still too large: if we plot $\delta$ versus \sp, we find that
while changing $a_1^1$ from its central value of 0.037 to a new
value of 0.087 changes $\delta$ essentially uniformly by 15
percent, changing instead $b_0^0$ from its central value of 0.24
to 0.19 has the same effect, to within about 1 percent.
Moreover,  with appropriate variations in $a_0^0$ and $b_1^1$,
one can quite well mock up a variation in $a_1^1$ (or $b_0^0$,
naturally).
Thus, one can only hope to determine two independent parameters
without recourse to other experiments or other theoretical constraints.

The MLM quantifies these conclusions.  Assuming Schenk's central
values ($a_0^0=0.2$, $b_1^1=0.005$, and above), $a_0^0$ or $b_1^1$
can be determined to 0.03 (for Rosselet statistics of 30,000 events),
or $b_0^0$ or $a_1^1$ to 0.04, if only one parameter is considered, and
the rest held fixed.  If one determines $a_0^0$ and $b_1^1$ simultaneously,
holding the rest of the parameters fixed,
the correlation is 0.71 (defined as the covariance of the two parameters
divided by the $\sigma$'s
for each, 1 for a maximally correlated pair of
variables), and the errors are $0.04$ and $0.05$ respectively.
If one determines instead $a^0_0$ and $b_0^0 - a^1_1$ simultaneously,
the correlation is $-0.9$ and the errors are $0.06$ and $0.08$;
for $b^1_1$ and $b_0^0 - a^1_1$ the correlation is $0.94$ and the
errors are $0.09$ and $0.11$.  To determine more than two parameters
simultaneously is not possible.

\chapter{CONCLUSION}
We have considered asymmetries and the
maximum likelihood method, and shown how the latter is more powerful
and yields more precise determinations.  We have found that the
current measurements of these parameters will be improved by a factor
of five to ten in one effective year ($10^7$ seconds) of running
with ${\cal L}= 5 \times 10^{32}
{\rm cm}^2 {\rm s}^{-1}$;
in particular we expect that the $\pi\pi$ scattering length
$a^0_0$ should be measureable to an accuracy of about $0.01$.

\chapter{ACKNOWLEDGEMENTS}
We would like to thank Gilberto Colangelo, Paolo Franzini and
Juliet Lee-Franzini for many useful discussions.

\par \penalty-400 \vskip\chapterskip
   \spacecheck\referenceminspace \immediate\closeout\referencewrite
   \referenceopenfalse
   \line{\fourteenrm\hfil REFERENCES\hfil}\vskip\headskip
   \input referenc.texauxil
   
\bye